\documentclass[aps,showpacs,preprintnumbers,amsmath, amssymb]{revtex4}

\usepackage{float}
\usepackage{graphics,epsfig}
\usepackage{graphicx}
\usepackage{dcolumn}
\usepackage{bm}

\begin{document}
\baselineskip=0.8 cm
\title{{\bf Holographic superconductor models with the Maxwell field strength corrections}}

\author{Qiyuan Pan$^{1,2}$\footnote{panqiyuan@126.com}, Jiliang Jing$^{1,2}$\footnote{jljing@hunnu.edu.cn}, Bin Wang$^{3}$\footnote{wang~b@sjtu.edu.cn}}
\affiliation{$^{1}$Institute of Physics and Department of Physics,
Hunan Normal University, Changsha, Hunan 410081, China}
\affiliation{$^{2}$ Key Laboratory of Low Dimensional Quantum
Structures and Quantum Control of Ministry of Education, Hunan
Normal University, Changsha, Hunan 410081, China}
\affiliation{$^{3}$ INPAC and Department of Physics, Shanghai Jiao
Tong University, Shanghai 200240, China}

\vspace*{0.2cm}
\begin{abstract}
\baselineskip=0.6 cm
\begin{center}
{\bf Abstract}
\end{center}

We study the effect of the quadratic field strength correction to
the usual Maxwell field on the holographic dual models
in the backgrounds of AdS black hole and AdS soliton. We
find that in the black hole background, the higher
correction to the Maxwell field makes the condensation harder to form and changes
the expected relation in the gap frequency. This effect is similar to that caused by the curvature correction. However, in the soliton background we find that different from the curvature effect, the correction to the Maxwell field does not influence the
holographic superconductor and insulator phase transition.

\end{abstract}


\pacs{11.25.Tq, 04.70.Bw, 74.20.-z}\maketitle
\newpage
\vspace*{0.2cm}

\section{Introduction}

As the most remarkable discovery in string theory, the anti-de
Sitter/conformal field theory (AdS/CFT) correspondence states that a
string theory on asymptotically AdS spacetimes can be related to a
conformal field theory on the boundary
\cite{Maldacena,Gubser1998,Witten}. Recently, this principle has
been employed to study the strongly correlated condensed matter
physics from the gravitational dual (for reviews, see
\cite{HartnollRev,HerzogRev,HorowitzRev}). It was shown that the
instability of the bulk black hole corresponds to a second order
phase transition from normal state to superconducting state which
brings the spontaneous U(1) symmetry breaking \cite{GubserPRD78}.
Due to the potential applications to the condensed matter physics,
the gravity models with the property of the so-called holographic
superconductor have been studied extensively, see for example
\cite{HartnollPRL101,HartnollJHEP12,HorowitzPRD78,
Nakano-Wen,Koutsoumbas,BasuPRD79,Maeda79,Sonner,Gubser-C-S-T,Gauntlett-Sonner,Cai-Zhang,
Chen-Jing,Jing-Chen,Ammon-et,Konoplya,Franco,Aprile-Russo,Herzog-2010,
Siopsis,Chen-Liu,Setare-Momeni,Wu-Cao} and references therein.

Recently, motivated by the application of the Mermin-Wagner theorem
to the holographic superconductors, there have been a lot of  interest in
exploring the effect of the curvature correction on the
($3+1$)-dimensional superconductor \cite{Gregory} and higher
dimensional ones \cite{Pan-Wang} by examining the charged scalar
field together with a Maxwell field in the Gauss-Bonnet-AdS black
hole background
\begin{eqnarray}\label{BH action}
S=\int d^{d}x\sqrt{-g}\left\{\frac{1}{16\pi
G}\left[R+\frac{(d-1)(d-2)}{L^2}+\mathcal{L}_{R^{2}}\right]
+\left(-\frac{1}{4}F_{\mu\nu}F^{\mu\nu}-|\nabla\psi - iA\psi|^{2}
-m^2|\psi|^2 \right)\right\} \;,\nonumber\\
\end{eqnarray}
with the curvature correction reads
\begin{eqnarray}\label{GB-correction}
\mathcal{L}_{R^{2}}=\tilde{\alpha}
\left(R_{\mu\nu\gamma\delta}R^{\mu\nu\gamma\delta}-4R_{\mu\nu}R^{\mu\nu}+
R^{2}\right),
\end{eqnarray}
where $\tilde{\alpha}$ is the Gauss-Bonnet coupling constant with
dimension $(length)^{2}$. It was observed that the higher curvature
correction makes the condensation harder to form and causes the behavior
of the claimed universal ratio $\omega/T_c\approx8$ unstable
\cite{Gregory,Pan-Wang,Brihaye,Ge-Wang,
Liu-Wang,BarclayGregory,Pan-WangPLB,Cai-Nie-Zhang,Kuang-Ling,Siani,
KannoCQG,JingGBI,Li-Cai-Zhang,Ge}.

As a matter of fact, in the low-energy limit of heterotic string
theory, the higher-order correction term appears also in the Maxwell
gauge field \cite{Gross-Sloan}. Thus, in order to understand the
influences of the $1/N$ or $1/\lambda$ ($\lambda$ is the 't Hooft
coupling) corrections on the holographic superconductors
\cite{Gregory}, it is interesting to consider the high-order
correction related to the gauge field besides the curvature
correction to the gravity. In this work, in order to grasp the
influence of the correction to the gauge field,  we will turn off
the curvature correction and study in a pure Einstein gravity
background for simplicity. We will consider a gauge field and the
scalar field coupled via a generalized Lagrangian
\begin{eqnarray}\label{System}
S=\int d^{d}x\sqrt{-g}\left\{\frac{1}{16\pi
G}\left[R+\frac{(d-1)(d-2)}{L^2}\right]
+\left(-\frac{1}{4}F_{\mu\nu}F^{\mu\nu}+\mathcal{L}_{F^{4}}
-|\nabla\psi-iA\psi|^{2}-m^2|\psi|^2\right)\right\} \ ,\nonumber\\
\end{eqnarray}
where $\mathcal{L}_{F^{4}}$ is determined by a quadratic field
strength correction to the usual Einstein-Maxwell field
\cite{Kats-Motl,Anninos,Liu-Szepietowski,Maeda-Hassaine,Cai-Pang}
\begin{eqnarray}
\mathcal{L}_{F^{4}}=c_{1}(F_{\mu\nu}F^{\mu\nu})^{2}
+c_{2}F_{\mu\nu}F^{\nu\delta}F_{\delta\kappa}F^{\kappa\mu},
\end{eqnarray}
with the real numbers $c_{1}$ and $c_{2}$. When $c_{1}$ and $c_{2}$
are zero, it reduces to the models considered in
\cite{HartnollPRL101,HartnollJHEP12,HorowitzPRD78}. Interestingly,
just as shown in the following discussion, the constraint can be
relaxed to the case $2c_{1}+c_{2}=0$ where the model (\ref{System})
reduces to the standard holographic superconductors studied in
\cite{HartnollPRL101,HartnollJHEP12,HorowitzPRD78}.

Recently the solutions of electrically charged black hole with the
higher correction term in the Maxwell field have been discussed
widely \cite{Kats-Motl,Anninos,Liu-Szepietowski,Maeda-Hassaine}, it
is interesting to study the coupling of the scalar field with the
higher order corrected Maxwell field, explore the effect of the
higher correction in the Maxwell field on the scalar condensation
and compare with the effect of the curvature correction.

Besides the black hole background, we will also extend our discussion to the AdS soliton background.  There have been a lot of work discussing the
 holographic
insulator and superconductor phase transitions in the
five-dimensional AdS soliton background
\cite{Nishioka-Ryu-Takayanagi,Horowitz-Soliton,Akhavan-Soliton,Basu-Soliton,brihaye-Soliton,
Cai-Li-Zhang,Peng-Pan-Wang-Soliton,Pan-Jing-Wang-Soliton}. The
discussion with the curvature correction in the Ricci flat AdS
soliton in Gauss-Bonnet gravity was discussed in
\cite{Pan-Wang,Pan-Jing-Wang-Soliton}. It would be of great interest
to examine the influence of the correction to the Maxwell field on
the holographic insulator and superconductor system. In this work,
we will compare the correction to the gauge field with the
correction to the curvature on the condensation in the  AdS soliton
background.

In order to extract the main physics, in this work we will concentrate on the probe limit to avoid the complex computation.  The organization of
the work is as follows. In Sec. II, we will study the holographic
superconductor models with $F^{4}$ corrections in the
Schwarzschild-AdS black hole background. In Sec. III we will extend our discussion to the Schwarzschild-AdS soliton background. We will
conclude in the last section of our main results.

\section{Holographic superconducting models with $F^{4}$ corrections}

We consider the background of the $d$-dimensional planar
Schwarzschild-AdS black hole
\begin{eqnarray}\label{BH metric}
ds^2=-f(r)dt^{2}+\frac{dr^2}{f(r)}+r^{2}dx_{i}dx^{i},
\end{eqnarray}
where
\begin{eqnarray}
f(r)=\frac{r^2}{L^2}\left(1-\frac{r_{+}^{d-1}}{r^{d-1}}\right),
\end{eqnarray}
 $L$ is the AdS radius and $r_{+}$ is the black hole horizon. The  Hawking temperature can be expressed as
\begin{eqnarray}
T=\frac{(d-1)r_{+}}{4\pi L^2},
\end{eqnarray}
which can be interpreted as the temperature of the CFT.

Taking the ansatz $\psi=|\psi|$, $A_{t}=\phi$ where $\psi$, $\phi$
are both real functions of $r$ only, we can obtain the equations of
motion from the action (\ref{System}) in the probe limit
\begin{eqnarray}
&&\psi^{\prime\prime}+\left(
\frac{d-2}{r}+\frac{f^\prime}{f}\right)\psi^\prime
+\left(\frac{\phi^2}{f^2}-\frac{m^2}{f}\right)\psi=0\,,
\label{BHPsi}
\end{eqnarray}
\begin{eqnarray}\left(1+3\varepsilon\phi^{\prime
2}\right)\phi^{\prime\prime}+\frac{d-2}{r}\left(1+\varepsilon\phi^{\prime
2}\right)\phi^\prime-\frac{2\psi^{2}}{f}\phi=0~, \label{BHPhi}
\end{eqnarray}
where we have set $\varepsilon=8(2c_{1}+c_{2})$ which can be used to
describe the $F^{4}$ correction to the usual Maxwell
field. Obviously, Eqs. (\ref{BHPsi}) and  (\ref{BHPhi})
reduce to the standard holographic superconductor models discussed
in \cite{HartnollPRL101,HartnollJHEP12,HorowitzPRD78} when
$\varepsilon=0$.

The equations of motion (\ref{BHPsi}) and  (\ref{BHPhi}) can be
solved numerically by doing integration from the horizon out to the
infinity. At the horizon $r=r_{+}$, the regularity gives the
boundary conditions
\begin{eqnarray}
\psi(r_{+})=\frac{f^\prime(r_{+})}{m^{2}}\psi^\prime(r_{+})\,,\hspace{0.5cm}
\phi(r_{+})=0\,. \label{horizon}
\end{eqnarray}
At the asymptotic AdS boundary $r\rightarrow\infty$, the solutions
behave like
\begin{eqnarray}
\psi=\frac{\psi_{-}}{r^{\lambda_{-}}}+\frac{\psi_{+}}{r^{\lambda_{+}}}\,,\hspace{0.5cm}
\phi=\mu-\frac{\rho}{r^{d-3}}\,, \label{infinity}
\end{eqnarray}
where $\mu$ and $\rho$ are interpreted as the chemical potential and
charge density in the dual field theory respectively, and
$\lambda_\pm=\frac{1}{2}[(d-1)\pm\sqrt{(d-1)^{2}+4m^{2}L^2}~]$. It
should be noted that the coefficients $\psi_{-}$ and $\psi_{+}$ both
multiply normalizable modes of the scalar field equations and they
correspond to the vacuum expectation values
$<\mathcal{O}_{-}>=\psi_{-}$, $<\mathcal{O}_{+}>=\psi_{+}$ of an
operator $\mathcal{O}$ dual to the scalar field according to the
AdS/CFT correspondence. Just as in Refs.
\cite{HartnollPRL101,HartnollJHEP12}, we can impose boundary
condition that either $\psi_{+}$ or $\psi_{-}$ vanishes.

\subsection{The condensation of the scalar operators}

In order to discuss the effects of the $F^{4}$ correction terms
$\varepsilon$ on the condensation of the scalar operators, we will
solve the equations of motion (\ref{BHPsi}) and  (\ref{BHPhi})
numerically. Since we focus on the effects of the $F^{4}$
corrections, we will set $d=4$ and $m^2L^2=-2$ for concreteness. As
a matter of fact, the other choices of the dimensionality of the
spacetime and the mass of the scalar field will not qualitatively
modify our results. It should be noted that, unlike the Gauss-Bonnet
holographic superconductors which should be in $3+1$ dimensions at
least, we can even construct $(2+1)$-dimensional holographic
superconducting models with $F^{4}$ corrections.

\begin{figure}[ht]
\includegraphics[scale=0.75]{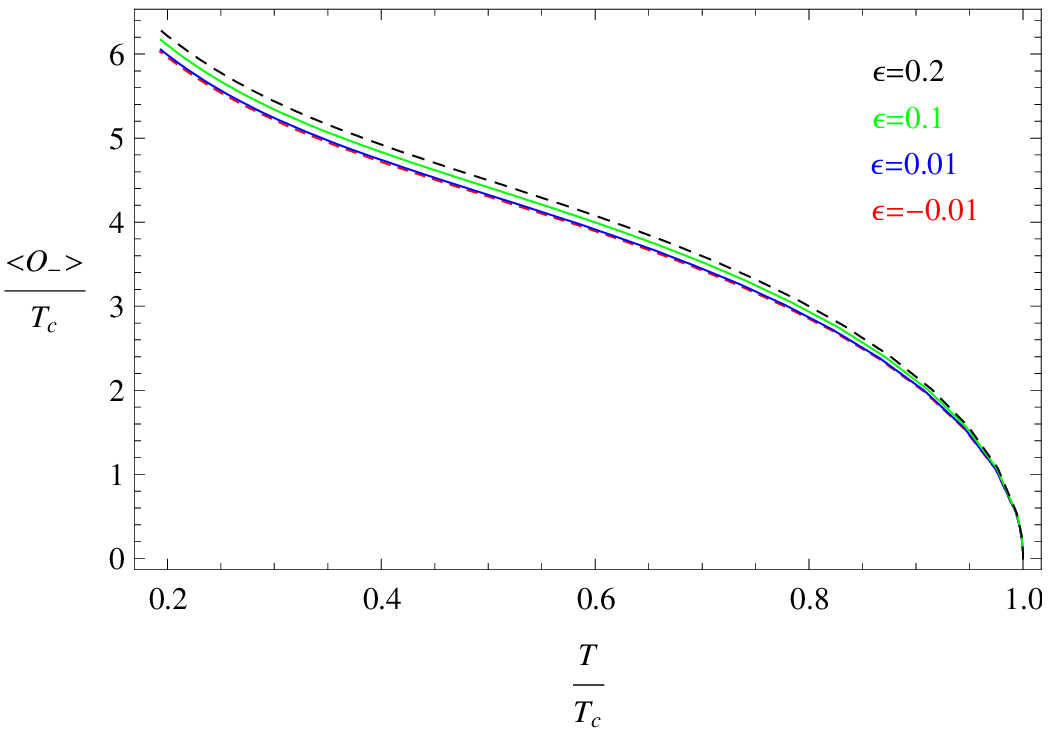}\vspace{0.0cm}
\includegraphics[scale=0.75]{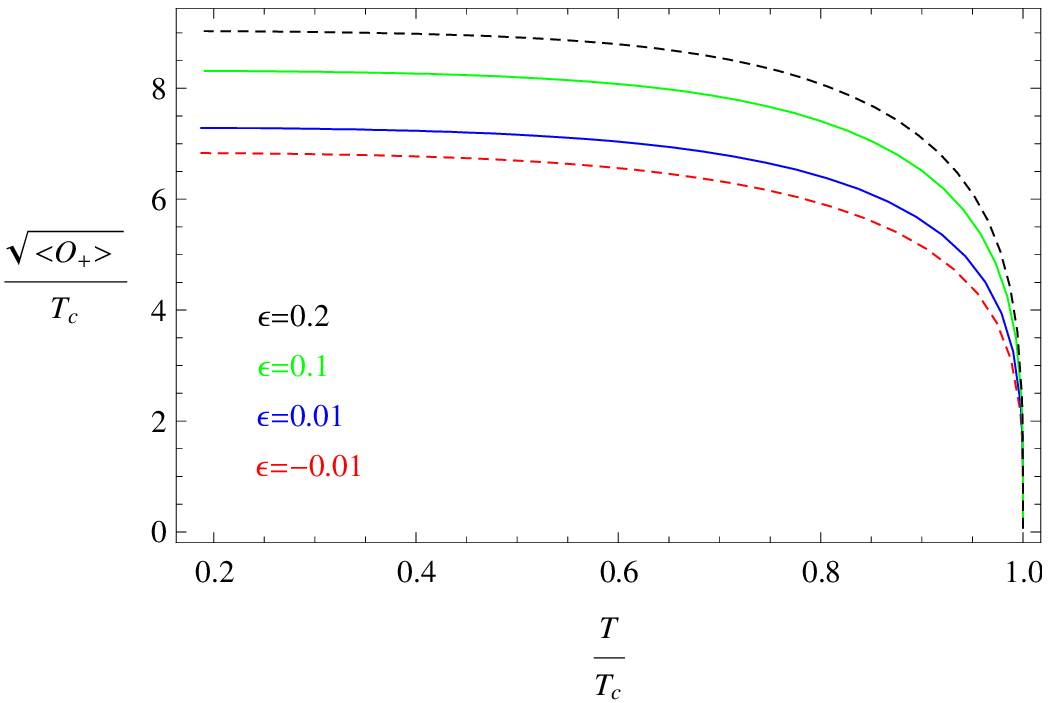}\\ \vspace{0.0cm}
\caption{\label{CondCMF} (color online) The condensates of the
scalar operators $\mathcal{O}_{-}$ and $\mathcal{O}_{+}$ as a
function of temperature for the mass of the scalar field $m^2L^2=-2$
in $d=4$ dimension. The four lines from bottom to top correspond to
increasing correction term, i.e., $\varepsilon=-0.01$ (red and
dashed), $0.01$ (blue), $0.1$ (green) and $0.2$ (black and dashed)
respectively. }
\end{figure}

In Fig. \ref{CondCMF} we present the condensates of the scalar
operators $\mathcal{O}_{-}$ and $\mathcal{O}_{+}$ as a function of
temperature with various correction terms $\varepsilon$ for the mass
of the scalar field $m^2L^2=-2$ in $d=4$ dimension. Obviously, the
curves in the right panel of Fig. \ref{CondCMF} have similar
behavior to the BCS theory for different $\varepsilon$, where the
condensate goes to a constant at zero temperature. However, the
curves for the operator $\mathcal{O}_{-}$ will diverge at low
temperature, which are similar to that for the usual Maxwell
electrodynamics in the probe limit neglecting backreaction of the
spacetime \cite{HartnollPRL101}. The behaviors of the condensates
for the scalar operators $\mathcal{O}_{-}$ and $\mathcal{O}_{+}$
show that the holographic superconductors still exist even we
consider $F^{4}$ correction terms to the usual Maxwell
electrodynamics.

\begin{table}[ht]
\caption{\label{Tc-D4} The critical temperature $T_{c}$ for the
operators $\mathcal{O}_{-}$ and $\mathcal{O}_{+}$ with different
values of $\varepsilon$ for $d=4$ and $m^2L^2=-2$. We have set
$\rho=1$ in the table.}
\begin{tabular}{c c c c c c c c c}
         \hline
$\varepsilon$ & -0.01 & -0.001 & 0 & 0.001 & 0.01 & 0.1 & 0.2 & 0.3
        \\
        \hline
~~~$\mathcal{O}_{-}$~~~&~~~~$0.2260$~~~~&~~~~$0.2256$~~~~&~~~~$0.2255(4)$~~~~&~~~~$0.2255(0)$~~~~&
~~~~$0.2251$~~~~&~~~~$0.2219$~~~&~~~~$0.2189$~~~&~~~~$0.2163$~~~
          \\
~~~$\mathcal{O}_{+}$~~~&~~~~$0.1233$~~~~&~~~~$0.1188$~~~~&~~~~$0.1184$~~~~&~~~~$0.1181$~~~~&
~~~~$0.1151$~~~~&~~~~$0.0993$~~~&~~~~$0.0900$~~~&~~~~$0.0836$~~~
          \\
        \hline
\end{tabular}
\end{table}

From Fig. \ref{CondCMF}, we see the higher correction term
$\varepsilon$ makes the condensation gap larger for both scalar
operators $\mathcal{O}_{-}$ and $\mathcal{O}_{+}$, which means that
the scalar hair is harder to be formed when adding $F^{4}$
corrections to the usual Maxwell field. In
fact, the table \ref{Tc-D4} shows that the critical temperature
$T_{c}$ for the operators $\mathcal{O}_{-}$ and $\mathcal{O}_{+}$
decreases as the correction term $\varepsilon$ increases, which
agrees well with the finding in Fig. \ref{CondCMF}. This behavior is
reminiscent of that seen for the Gauss-Bonnet holographic
superconductors, where the higher curvature corrections make
condensation harder, so we conclude that the $F^{4}$ corrections to
the usual Maxwell field and the curvature corrections
share some similar features for the condensation of the scalar
operators.

\subsection{Conductivity}

Now we are in a position to investigate the influence of the $F^{4}$
correction term on the conductivity. Since the condensation gap and
the critical temperature depend on the correction term $\varepsilon$
which is similar to the Gauss-Bonnet correction term in the
holographic superconductor, we want to know whether the
correction term $\varepsilon$ will change the expected universal
relation $\omega_g/T_c\approx 8$ in the gap frequency
\cite{HorowitzPRD78}  as the Gauss-Bonnet term did.

Considering the perturbed Maxwell field $\delta
A_{x}=A_{x}(r)e^{-i\omega t}dx$, we obtain the equation of motion
for $\delta A_{x}$ which can be used to calculate the conductivity
\begin{eqnarray}
A_{x}^{\prime\prime}+\left(\frac{d-4}{r}+\frac{f^\prime}{f}+
\frac{2\varepsilon\phi^{\prime}\phi^{\prime\prime}}{1+\varepsilon\phi^{\prime
2}}\right)A_{x}^\prime
+\left[\frac{\omega^2}{f^2}-\frac{2\psi^{2}}{f(1+\varepsilon\phi^{\prime
2})}\right]A_{x}=0 \; . \label{Conductivity Equation}
\end{eqnarray}
We still restrict our study to $d=4$ for simplicity. Though the
above equation is more complicated than that in usual
Einstein-Maxwell electrodynamics, the ingoing wave boundary
condition near the horizon is still given by
\begin{eqnarray}
A_{x}(r)\sim f(r)^{-\frac{i\omega}{3r_+}},
\end{eqnarray}
and in the asymptotic AdS region
\begin{eqnarray}
A_{x}=A^{(0)}+\frac{A^{(1)}}{r}.
\end{eqnarray}
Thus, we can obtain the conductivity of the dual superconductor by
using the AdS/CFT dictionary \cite{HartnollPRL101,HartnollJHEP12}
\begin{eqnarray}\label{CMFConductivity}
\sigma=-\frac{iA^{(1)}}{\omega A^{(0)}}\ .
\end{eqnarray}
For different values of $F^{4}$ correction term $\varepsilon$, one
can obtain the conductivity by solving the Maxwell equation
numerically. We will focus on the case for the fixed scalar mass
$m^2L^2=-2$ in our discussion.

\begin{figure}[ht]
\includegraphics[scale=0.51]{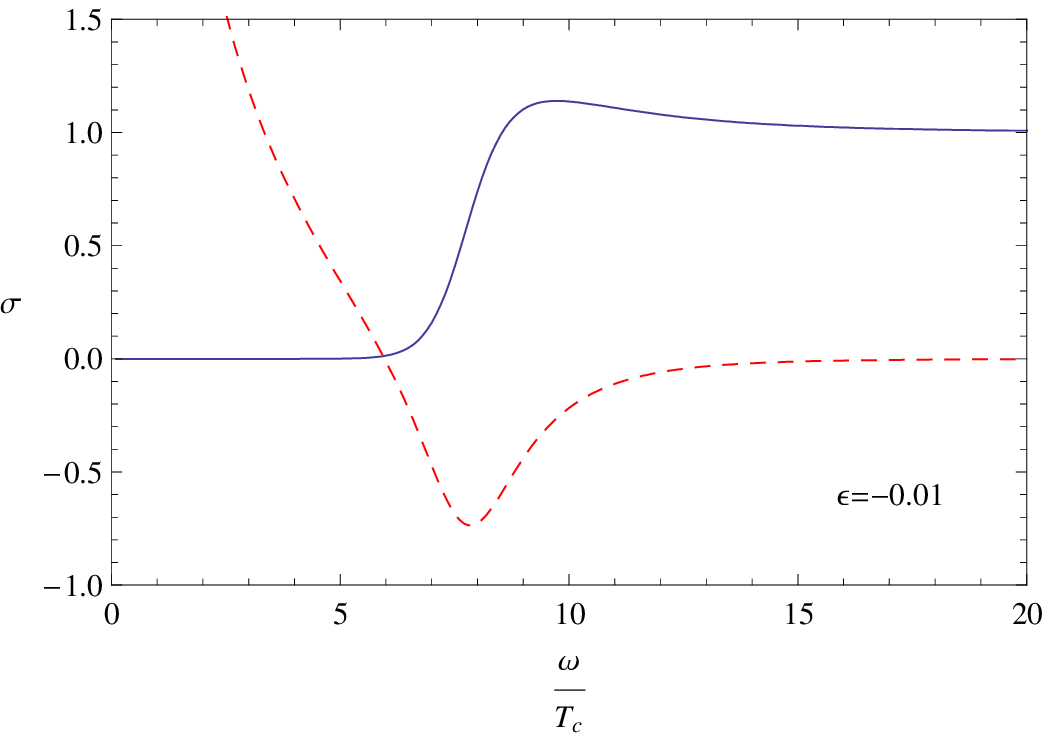}\hspace{0.2cm}%
\includegraphics[scale=0.51]{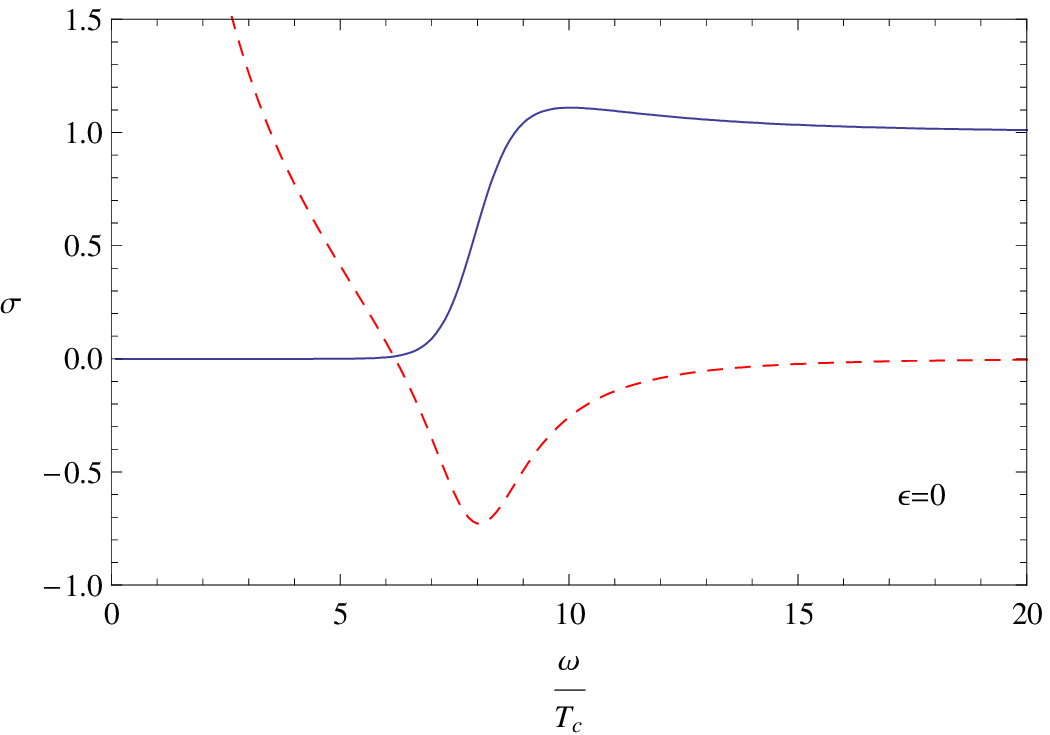}\vspace{0.0cm}
\includegraphics[scale=0.51]{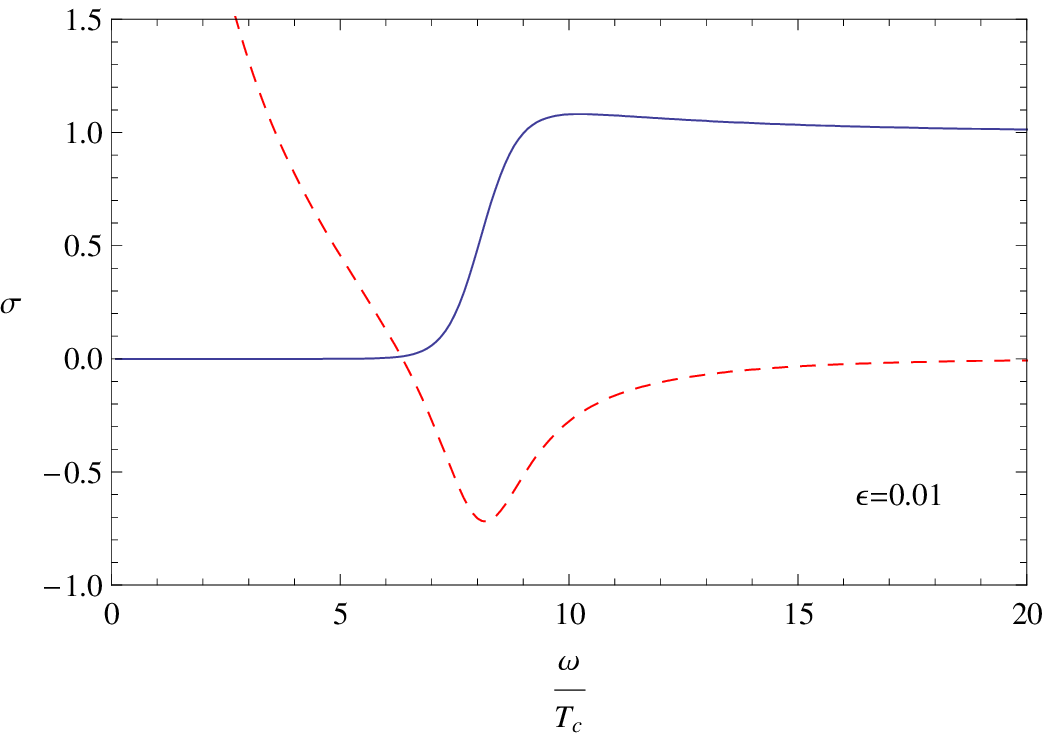}\\ \hspace{0.2cm}%
\includegraphics[scale=0.51]{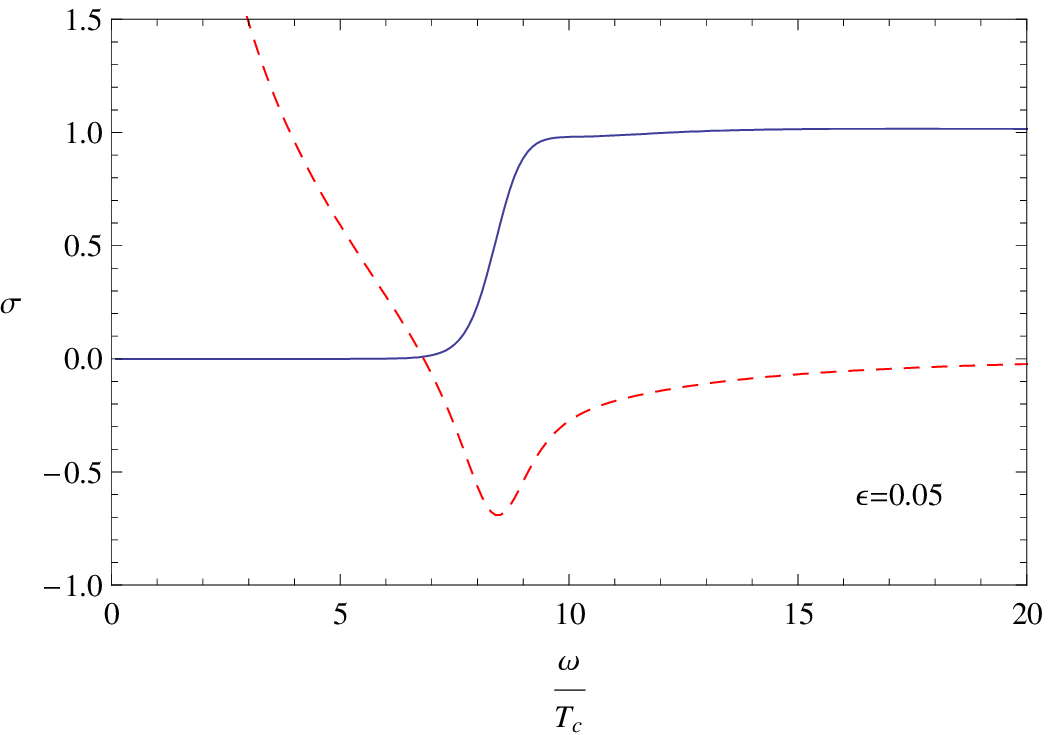}\vspace{0.0cm}
\includegraphics[scale=0.51]{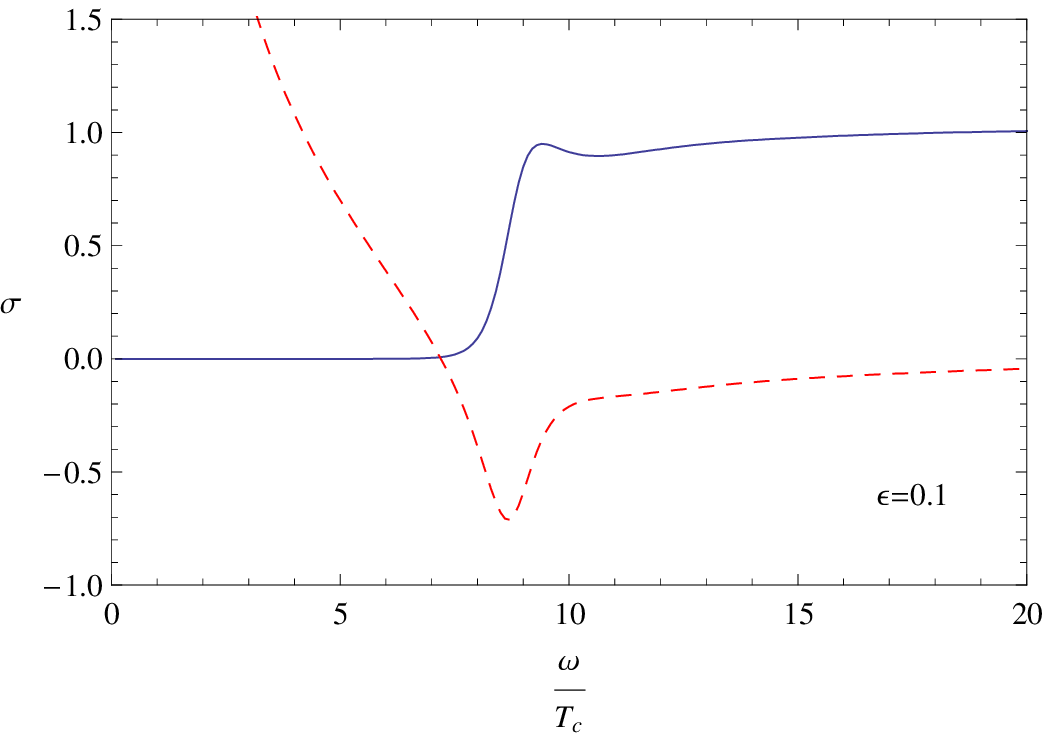}\hspace{0.2cm}%
\includegraphics[scale=0.51]{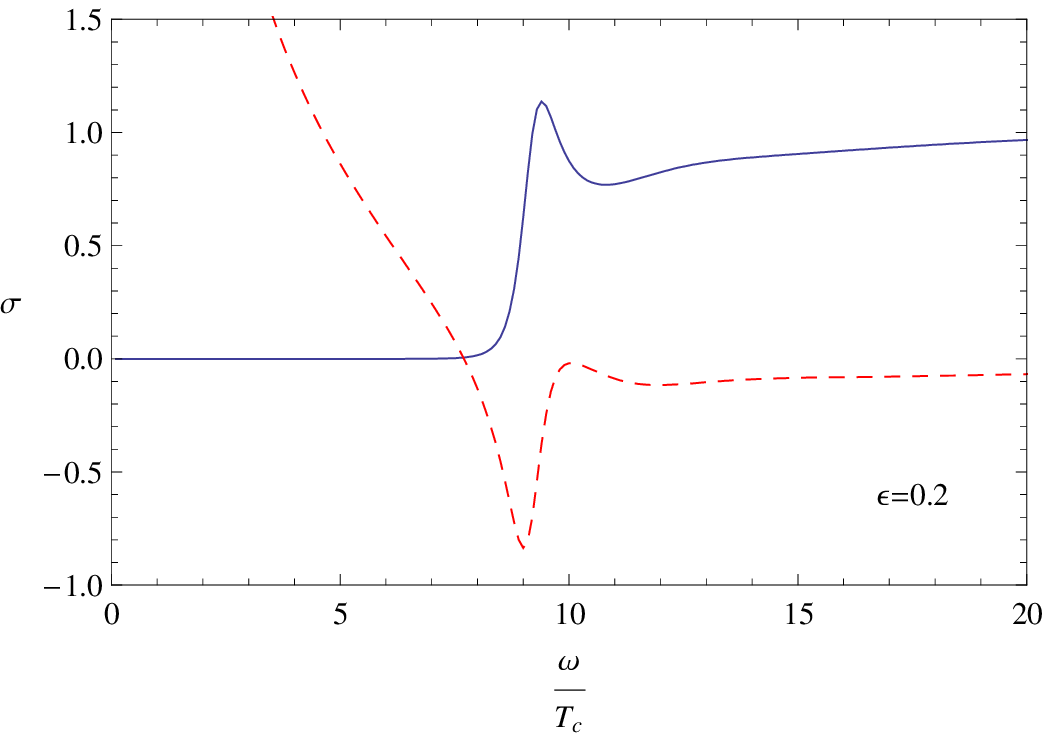}\\ \vspace{0.0cm}
\caption{\label{CMFConductivity} (color online) Conductivity of
($2+1$)-dimensional superconductors with the $F^{4}$ corrections for
the fixed mass of the scalar field $m^2L^2=-2$ and different
correction terms, i.e., $\varepsilon=-0.01$, $0$, $0.01$, $0.05$,
$0.1$ and $0.2$.}
\end{figure}

In Fig. \ref{CMFConductivity} we plot the frequency dependent
conductivity obtained by solving the Maxwell equation
(\ref{Conductivity Equation}) numerically for $\varepsilon=-0.01$,
$0$, $0.01$, $0.05$, $0.1$ and $0.2$ at temperatures
$T/T_{c}\approx0.2$. The blue (solid) line and red (dashed) line
represent the real part and imaginary part of the conductivity
$\sigma(\omega)$ respectively. We find a gap in the conductivity
with the gap frequency $\omega_{g}$. For the same mass of the scalar
field, we observe that with the increase of the $F^{4}$ correction
term $\varepsilon$, the gap frequency $\omega_{g}$ becomes larger.
Also, for increasing $F^{4}$ correction term, we have larger
deviations from the value $\omega_g/T_c\approx 8$. This shows that
the high $F^{4}$ corrections really change the expected universal
relation in the gap frequency, which is similar to the effect of the
Gauss-Bonnet coupling.

\section{Holographic superconductor/insulator transitions with $F^{4}$ corrections}

In Refs. \cite{Pan-Wang,Pan-Jing-Wang-Soliton}, we discussed the
holographic dual to  Gauss-Bonnet-AdS soliton in the usual Maxwell
electrodynamics. It shows that although the Gauss-Bonnet term has no
effect on the Hawking-Page phase transition between AdS black hole
and AdS soliton, it does have an effect on the scalar condensation
and conductivity in Gauss-Bonnet-AdS soliton configuration. In this
section we will examine the effect of $F^{4}$ correction in the Schwarzschild-AdS soliton
background and explore its influence on the insulator and superconductor phase transition.

\subsection{Superconductor/insulator phase in the AdS soliton}

Making use of two wick rotations for the AdS Schwarzschild black
hole given in (\ref{BH metric}), we can obtain the $d$-dimensional
AdS soliton
\begin{eqnarray}\label{soliton}
ds^2=-r^{2}dt^{2}+\frac{dr^2}{f(r)}+f(r)d\varphi^2+r^{2}dx_{j}dx^{j},
\end{eqnarray}
with
\begin{eqnarray}
f(r)=\frac{r^2}{L^2}\left(1-\frac{r_{s}^{d-1}}{r^{d-1}}\right).
\end{eqnarray}
Note that there does not exist any horizon in this solution and
$r=r_{s}$ is a conical singularity. Imposing a period
$\beta=\frac{4\pi L^{2}}{(d-1)r_{s}}$ for the coordinate $\varphi$,
we can remove the singularity.

Beginning with the generalized Lagrangian (\ref{System}), we can get
the equations of motion for the scalar field and gauge field in the
probe limit
\begin{eqnarray}
\psi^{\prime\prime}+\left(\frac{d-2}{r}
+\frac{f^\prime}{f}\right)\psi^\prime
+\left(\frac{\phi^2}{r^2f}-\frac{m^2}{f}\right)\psi=0\,,
\label{solitonPsi}
\end{eqnarray}
\begin{eqnarray}
\left(1+\frac{3\varepsilon f}{r^{2}}\phi^{\prime
2}\right)\phi^{\prime\prime}+\left[\frac{d-4}{r}+\frac{f^\prime}{f}+\varepsilon
\left(\frac{2f^\prime}{r^{2}}-\frac{f}{r^{3}}\right)\phi^{\prime
2}\right] \phi^\prime-\frac{2\psi^2}{f}\phi=0. \label{solitonPhi}
\end{eqnarray}
Using the shooting method, we will solve these two equations
numerically with appropriate boundary conditions at $r=r_{s}$ and at
the boundary $r\rightarrow\infty$. At the tip $r=r_{s}$, the
solutions behave like
\begin{eqnarray}
\psi=\tilde{\psi}_{0}+\tilde{\psi}_{1}(r-r_{s})+\tilde{\psi}_{2}(r-r_{s})^{2}+\cdots\,, \nonumber \\
\phi=\tilde{\phi}_{0}+\tilde{\phi}_{1}(r-r_{s})+\tilde{\phi}_{2}(r-r_{s})^{2}+\cdots\,,
\label{SolitonBoundary}
\end{eqnarray}
where $\tilde{\psi}_{i}$ and $\tilde{\phi}_{i}$ ($i=0,1,2,\cdots$)
are integration constants, and we impose the Neumann-like boundary
condition to keep every physical quantity finite
\cite{Nishioka-Ryu-Takayanagi}. Obviously, we can find a constant
nonzero gauge field $\phi(r_{s})$ at $r=r_{s}$. This is in strong
contrast to the AdS black hole, where $\phi(r_{+})=0$ at the
horizon. Near the AdS boundary $r\rightarrow\infty$, the solutions
have the same form just as Eq. (\ref{infinity}). For clarity, we
will take $d=5$ and still use the probe approximation in our
calculation.

It is well-known that the solution is unstable and a hair can be
developed when the chemical potential is bigger than a critical
value, i.e., $\mu>\mu_{c}$. However, the gravitational dual is an
AdS soliton with a nonvanishing profile for the scalar field $\psi$
if $\mu<\mu_{c}$, which can be viewed as an insulator phase
\cite{Nishioka-Ryu-Takayanagi}. Thus, around the critical chemical
potential $\mu_{c}$ there is a phase transition between the
insulator and superconductor phases. We will examine the effect of
the $F^{4}$ correction term on $\mu_{c}$ numerically.

\begin{figure}[ht]
\includegraphics[scale=0.75]{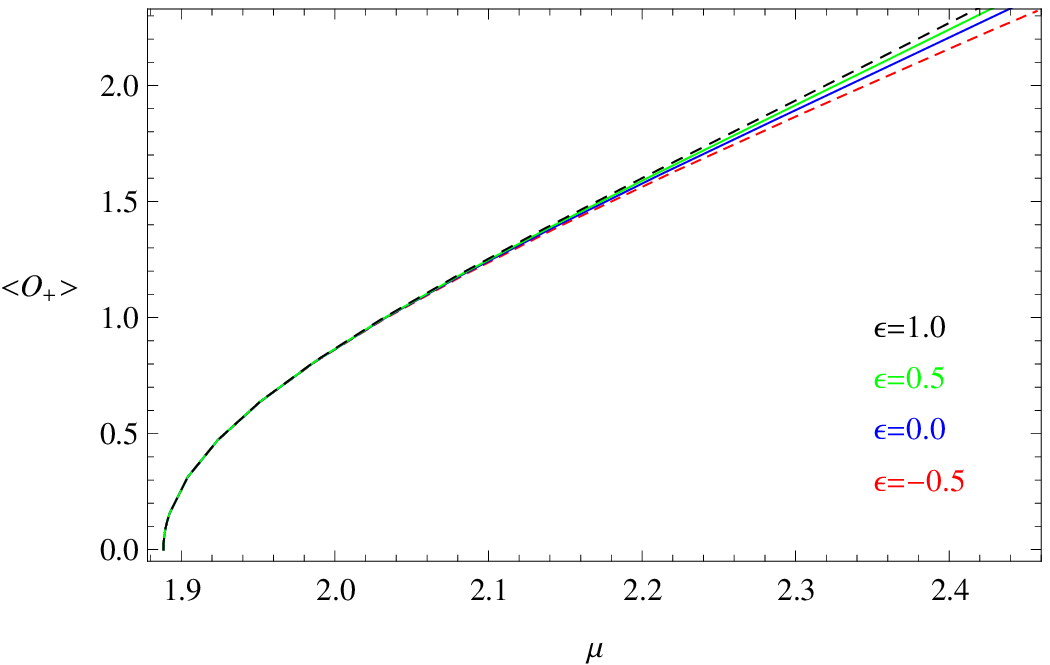}\vspace{0.0cm}
\includegraphics[scale=0.75]{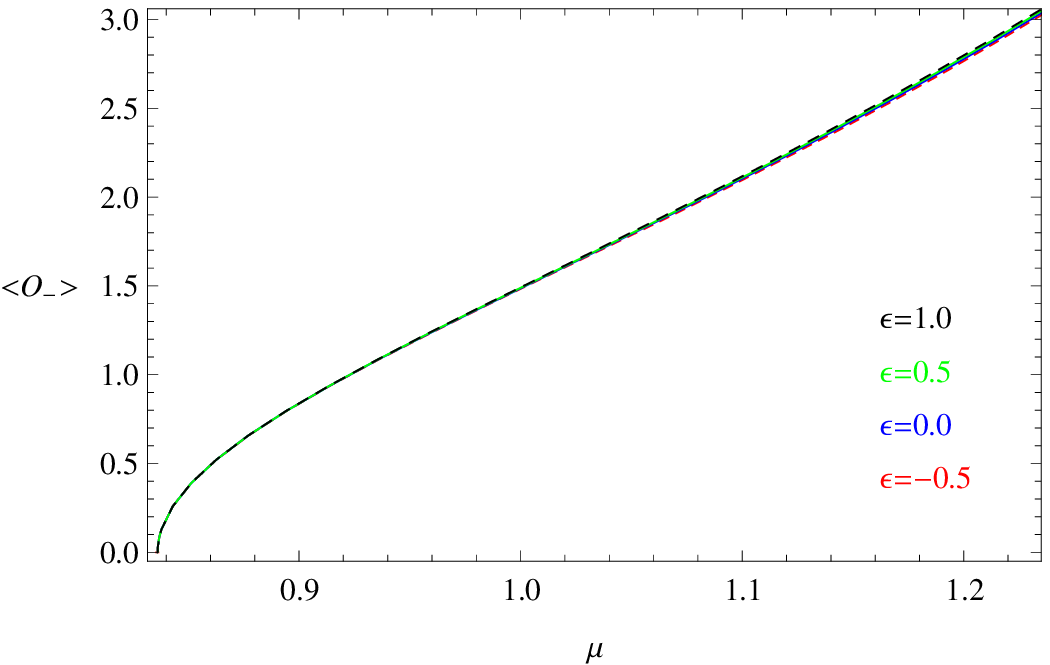}\\ \vspace{0.0cm}
\caption{\label{CondCMFsoliton} (color online) The condensates of
the scalar operators $\mathcal{O}_{+}$ and $\mathcal{O}_{-}$ with
respect to the chemical potential $\mu$ for the mass of the scalar
field $m^{2}L^{2}=-15/4$ in $d=5$ dimension. In each panel, the four
lines from bottom to top correspond to increasing correction term,
i.e., $\varepsilon=-0.5$ (red and dashed), $0.0$ (blue), $0.5$ (green)
and $1.0$ (black and dashed) respectively. }
\end{figure}

In Fig. \ref{CondCMFsoliton} we plot the condensations of scalar
operators $\cal O_{+}$ and $\cal O_{-}$ with respect to the chemical
potential $\mu$ in the $5$-dimensional AdS Soliton for different
$F^{4}$ correction terms with the fixed scalar mass
$m^{2}L^2=-15/4$. From this figure, we find that the critical
chemical potential $\mu_{c}$ is independent of the correction term
$\varepsilon$. As a matter of fact, selecting the mass of the scalar
field in the range
$-\frac{(d-1)^{2}}{4}<m^{2}L^2<-\frac{(d-1)^{2}}{4}+1$ for $d=5$
where both modes of the asymptotic values of the scalar fields are
normalizable, we obtain $\mu_{c-}$ and $\mu_{c+}$ for scalar
operators $\langle{\cal O_{-}}\rangle$ and $\langle{\cal
O_{+}}\rangle$ with different values of $m$ and $\varepsilon$
respectively
\begin{eqnarray}
&&\mu_{c-}=0.409~~{\rm and}~~\mu_{c+}=2.261,\quad {\rm
for}~~m^{2}L^{2}=-13/4~~{\rm and}~~\forall\varepsilon,\nonumber\\
&&\mu_{c-}=0.598~~{\rm and}~~\mu_{c+}=2.099,\quad {\rm
for}~~m^{2}L^{2}=-7/2~~{\rm and}~~\forall\varepsilon,\nonumber\\
&&\mu_{c-}=0.836~~{\rm and}~~\mu_{c+}=1.888,\quad {\rm
for}~~m^{2}L^{2}=-15/4~~{\rm and}~~\forall\varepsilon.
\label{SolitonCCP}
\end{eqnarray}
It is shown the critical chemical potentials $\mu_{c-}$ and
$\mu_{c+}$ are independent of the correction term $\varepsilon$ for
the fixed mass of the scalar field, which is in contrast to the case
of considering the Gauss-Bonnet correction term
\cite{Pan-Wang,Pan-Jing-Wang-Soliton}. However, the critical
chemical potentials $\mu_{c-}$ and $\mu_{c+}$ depend on the mass of
the scalar field, i.e., $\mu_{c-}$ for the scalar operator $\cal
O_{-}$ becomes smaller with the increase of the scalar field mass,
but larger scalar filed mass leads higher $\mu_{c+}$ for the scalar
operator $\cal O_{+}$, which is in agreement with the results in
Refs. \cite{Nishioka-Ryu-Takayanagi,Pan-Wang}.

\begin{figure}[ht]
\includegraphics[scale=0.75]{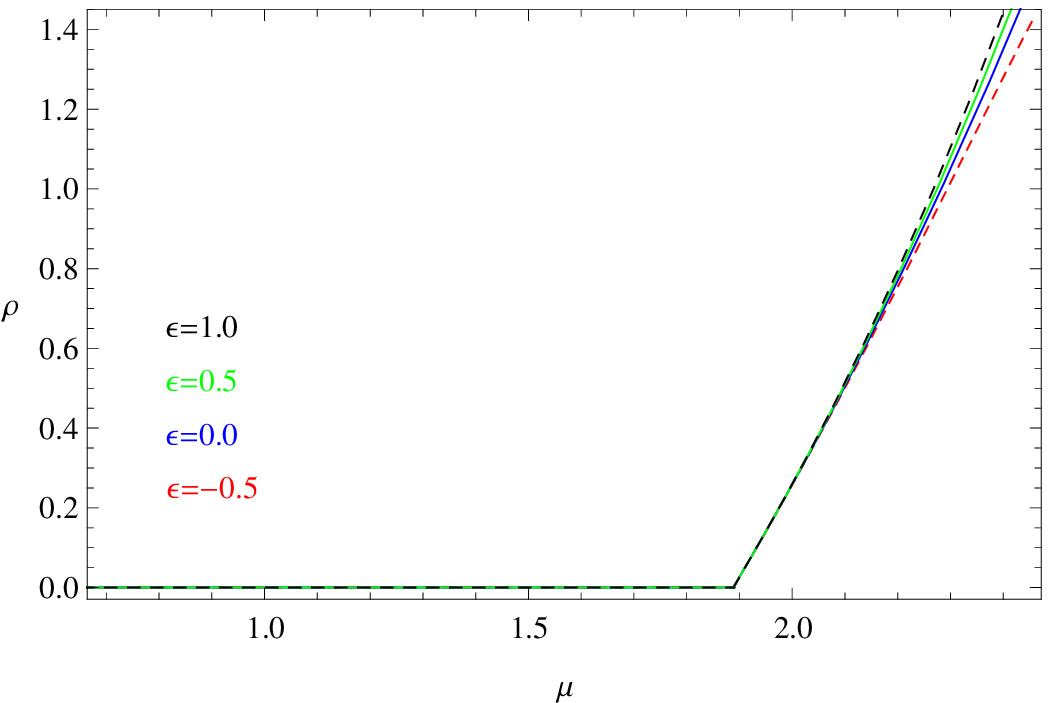}\vspace{0.0cm}
\includegraphics[scale=0.75]{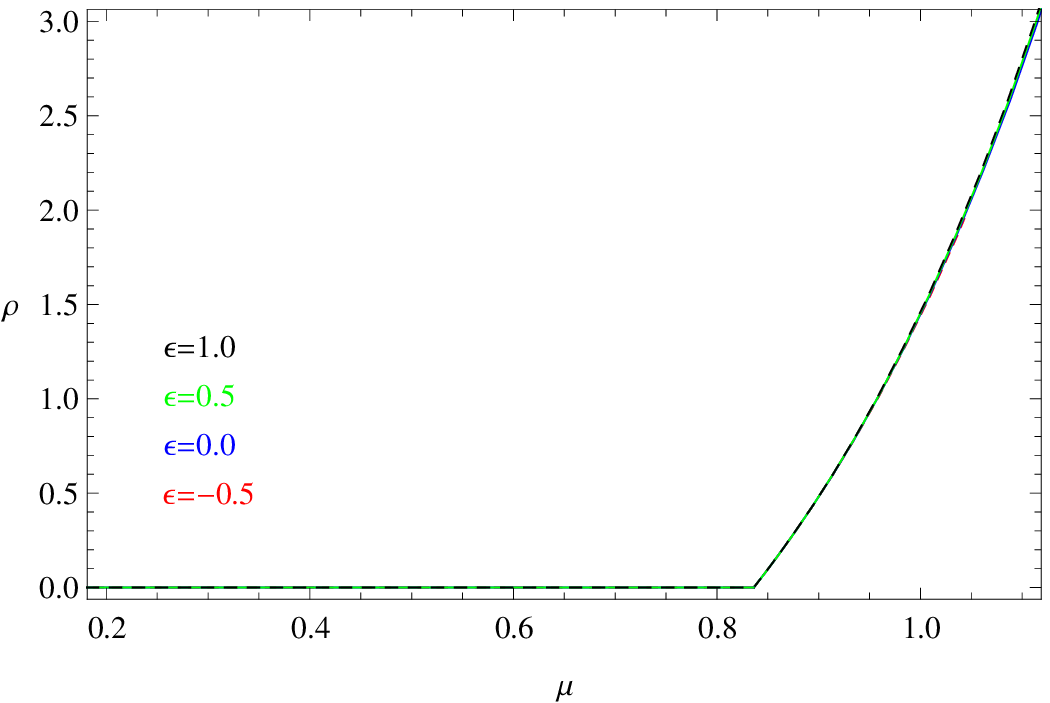}\\ \vspace{0.0cm}
\caption{\label{SLRealization} (color online) The charge density
$\rho$ as a function of the chemical potential $\mu$ with fixed mass
of the scalar field $m^{2}L^{2}=-15/4$ when $\langle{\cal
O_{+}}\rangle\neq0$ (left) and $\langle{\cal O_{-}}\rangle\neq0$
(right). Their derivatives jump at the phase transition points. In
each panel, the four lines from bottom to top correspond to
increasing correction term, i.e., $\varepsilon=-0.5$ (red and dashed),
$0.0$ (blue), $0.5$ (green) and $1.0$ (black and dashed)
respectively.}
\end{figure}

In Fig. \ref{SLRealization}, we plot the charge density $\rho$ as a
function of the chemical potential $\mu$ when $\langle{\cal
O_{+}}\rangle\neq0$ (left) and $\langle{\cal O_{-}}\rangle\neq0$
(right) for $m^{2}L^{2}=-15/4$. For each chosen $\varepsilon$, we
see that when $\mu$ is small, the system is described by the AdS
soliton solution itself, which can be interpreted as the insulator
phase \cite{Nishioka-Ryu-Takayanagi}. When $\mu$ reaches $\mu_{c-}$
or $\mu_{c+}$, there is a phase transition and the AdS soliton
reaches the superconductor (or superfluid) phase for larger $\mu$.
Still, we can find that the correction terms $\varepsilon$ do not
have any effect on the critical chemical potentials $\mu_{c-}$ and
$\mu_{c+}$ for the fixed mass of the scalar field.

\subsection{Analytical understanding of the superconductor/insulator phase transition}

Here we will apply the Sturm-Liouville method \cite{Siopsis} to
analytically investigate the properties of holographic
insulator/superconductor phase transition with $F^{4}$ corrections.
We will analytically calculate the critical chemical potential which
can accommodate the phase transition.

Introducing a new variable $z=1/r$, we can rewrite Eqs.
(\ref{solitonPsi}) and (\ref{solitonPhi}) for $d=5$ into
\begin{eqnarray}
\psi^{\prime\prime}+\left(
\frac{f^\prime}{f}-\frac{1}{z}\right)\psi^\prime
+\left(\frac{\phi^2}{z^2f}-\frac{m^2}{z^4f}\right)\psi=0\,,
\label{solitonPsi-Z}
\end{eqnarray}
\begin{eqnarray}
(1+3\varepsilon fz^{6}\phi^{\prime
2})\phi^{\prime\prime}+\left[\frac{1}{z}+\frac{f^\prime}{f}+\varepsilon
z^{5}(7f+2zf^\prime)\phi^{\prime 2}\right]
\phi^\prime-\frac{2\psi^2}{z^4f}\phi=0, \label{solitonPhi-Z}
\end{eqnarray}
where the prime denotes the derivative with respective to $z$.

At the critical chemical potential $\mu_{c}$, the scalar field
$\psi=0$. Thus, near the critical point Eq. (\ref{solitonPhi-Z})
reduces to
\begin{eqnarray}
(1+3\varepsilon fz^{6}\phi^{\prime
2})\phi^{\prime\prime}+\left[\frac{1}{z}+\frac{f^\prime}{f}+\varepsilon
z^{5}(7f+2zf^\prime)\phi^{\prime 2}\right] \phi^\prime=0.
\label{Phi-critical}
\end{eqnarray}
With the Neumann-like boundary condition (\ref{SolitonBoundary}) for
the gauge field $\phi$ at the tip $r=r_{s}$, we can obtain the
physical solution $\phi(z)=\mu$ to Eq. (\ref{Phi-critical}) when
$\mu<\mu_{c}$. Considering the asymptotic behavior given in Eq.
(\ref{infinity}), close to the critical point $\mu_{c}$, this
solution indicates that $\rho=0$ near the AdS boundary $z=0$, which
agrees with our previous numerical results.

As $\mu\rightarrow\mu_{c}$, the scalar field equation
(\ref{solitonPsi-Z}) reduces to
\begin{eqnarray}
\psi^{\prime\prime}+\left(\frac{f^\prime}{f}-\frac{1}{z}\right)\psi^\prime
+\left(\frac{\mu^2}{z^2f}-\frac{m^2}{z^4f}\right)\psi=0,
\label{criticalPsi}
\end{eqnarray}
which is the master equation to give the critical chemical potential
$\mu_{c}$ in the Sturm-Liouville method.

Before going further, we would like to comment Eq.
(\ref{criticalPsi}). Although Eq. (\ref{Phi-critical}) for $\phi$
depends on $\varepsilon$, but the correction terms $\varepsilon$ are
absent in the master Eq. (\ref{eigenvalue}). Thus, we can
immediately conclude that the $F^{4}$ correction term do not have
any effect on the critical chemical potential $\mu_{c}$ for the
fixed mass of the scalar field. However, for the black hole
background, due to the difference of boundary conditions at the
horizon, the physical solution $\phi(r)$ of Eq. (\ref{BHPhi})
depends on $\varepsilon$, this results in the appearance of the
correction term $\varepsilon$ in the master equation derived from
Eq. (\ref{BHPsi}). Thus, in this case the $F^{4}$ correction terms
do have effect on the critical temperature $T_{c}$ for the AdS black
hole, which agrees well to the numerical results.

Still, we will work on Eq. (\ref{criticalPsi}) to understand the
dependence of the critical chemical potential on the mass of the
scalar field analytically. As in \cite{Siopsis}, we introduce a
trial function $F(z)$ near the boundary $z=0$ which satisfies
\begin{eqnarray}\label{PhiFz}
\psi(z)\sim \langle{\cal O}_{i}\rangle z^{\lambda_i}F(z),
\end{eqnarray}
with $i=+$ or $i=-$. Note that the function $F(z)$ has the boundary
condition $F(0)=1$ and $F'(0)=0$ \cite{Siopsis}. So the equation of
motion for $F(z)$ is
\begin{eqnarray}\label{Fzmotion}
F^{\prime\prime}+\left[\frac{2\lambda_i}{z}+\left(\frac{f'}{f}-\frac{1}{z}\right)\right]
F^{\prime}+\left[\frac{\lambda_i(\lambda_{i}-1)}{z^2}+\frac{\lambda_i}{z}
\left(\frac{f'}{f}-\frac{1}{z}\right)+\frac{1}{z^{4}f}(\mu^{2}z^{2}-m^{2})\right]F=0.
\end{eqnarray}
Defining a new function
\begin{eqnarray}
T(z)=z^{2\lambda_{i}-3}(z^{4}-1),
\end{eqnarray}
we can rewrite Eq. (\ref{Fzmotion}) as
\begin{eqnarray}\label{NewFzmotion}
(TF^{\prime})^{\prime}+T\left[\frac{\lambda_i(\lambda_{i}-1)}{z^2}+\frac{\lambda_i}{z}
\left(\frac{f'}{f}-\frac{1}{z}\right)+\frac{1}{z^{4}f}(\mu^{2}z^{2}-m^{2})\right]F=0.
\end{eqnarray}
According to the Sturm-Liouville eigenvalue problem
\cite{Gelfand-Fomin}, we obtain the expression which will be used to
estimate the minimum eigenvalue of $\mu^2$
\begin{eqnarray}\label{eigenvalue}
\mu^{2}=\frac{\int^{1}_{0}T\left(F'^{2}-UF^{2}\right)dz}{\int^{1}_{0}VF^{2}dz},
\end{eqnarray}
with
\begin{eqnarray}
&&U=\frac{\lambda_{i}(\lambda_{i}-1)}{z^{2}}+\frac{\lambda_{i}}{z}\left(\frac{f'}{f}-\frac{1}{z}\right)
-\frac{m^{2}}{z^{4}f},\nonumber\\
&&V=\frac{T}{z^{2}f}.
\end{eqnarray}
In the following calculation, we will assume the trial function to
be $F(z)=1-az^{2}$, where $a$ is a constant.

For clarity, we will take $i=+$ and one can easily extend the study
to the case of $i=-$. From Eq. (\ref{eigenvalue}), we obtain the
expression for $i=+$
\begin{eqnarray}
\mu^{2}=\frac{\Sigma(a,\alpha)}{\Xi(a,\alpha)},
\end{eqnarray}
with
\begin{eqnarray}
\Sigma(a,m)&=&-\frac{4a^{2}}{12+m^{2}+6\sqrt{4+m^{2}}}-
\frac{8+m^{2}+4\sqrt{4+m^{2}}}{2}\left(\frac{1}{2+\sqrt{4+m^{2}}}-
\frac{2a}{3+\sqrt{4+m^{2}}}+\frac{a^{2}}{4+\sqrt{4+m^{2}}}\right),
\nonumber\\
\Xi(a,m)&=&-\frac{1}{2(1+\sqrt{4+m^{2}})}+\frac{a}{2+\sqrt{4+m^{2}}}-\frac{a^{2}}{2(3+\sqrt{4+m^{2}})}.
\end{eqnarray}
For different values of the mass of scalar field, we can get the
minimum eigenvalue of $\mu^{2}$ and the corresponding value of
$a$, for example, $\mu_{min}^{2}=5.121$ and $a=0.361$ for
$m^{2}L^{2}=-13/4$, $\mu_{min}^{2}=4.416$ and $a=0.348$ for
$m^{2}L^{2}=-14/4$, and $\mu_{min}^{2}=3.574$ and $a=0.330$ for
$m^{2}L^{2}=-15/4$. Then, we have the critical chemical potential
$\mu_{c}=\mu_{min}$ \cite{Cai-Li-Zhang}, i.e.,
\begin{eqnarray}
&&\mu_{c}=2.263,\quad {\rm
for}~~m^{2}L^{2}=-13/4,\nonumber\\
&&\mu_{c}=2.101,\quad {\rm
for}~~m^{2}L^{2}=-7/2,\nonumber\\
&&\mu_{c}=1.890,\quad {\rm for}~~m^{2}L^{2}=-15/4.
\end{eqnarray}
Comparing with numerical results in Eq. (\ref{SolitonCCP}), we find
that the analytic results derived from Sturm-Liouville method are in
good agreement with the numerical calculation.

Thus, we conclude that, unlike the holographic superconductor
models, the holographic superconductor/insulator transitions is not
affected by the $F^{4}$ correction terms but only depends on the
mass of the scalar field.

\section{Conclusions}

We have investigated the behavior of the holographic superconductors
in the presence of the a quadratic field strength correction $F^{4}$
to the usual Maxwell field. Different from the same order curvature
correction in the Gauss-Bonnet holographic dual models which only
appears in spacetime with dimension higher than $3+1$,  $F^{4}$
correction can appear basically in all dimensions. We found that
similar to the curvature correction, in the black hole background,
the higher $F^{4}$ correction term can make the condensation harder
to form and result in the larger deviations from the universal value
$\omega_g/T_c\approx 8$ for the gap frequency. Thus, the $F^{4}$
corrections and the Gauss-Bonnet corrections share some similar
features for the holographic superconductor system. However, the
story is completely different if we study the holographic
superconductor/insulator transitions with the $F^{4}$ correction. In
contrast to the curvature correction effect we observed in the
Gauss-Bonnet AdS soliton background that the critical chemical
potentials are independent of the $F^{4}$ correction term, which
tells us that the correction to the Maxwell field will not affect
the properties of the holographic superconductor/insulator phase
transition. We confirmed our numerical result by using the
Sturm-Liouville analytic method and concluded that different from
the AdS black hole background, the corrections to the gravity and
gauge field do play different roles in  the holographic
superconductor and insulator phase transition.

\begin{acknowledgments}

This work was supported by the National Natural Science Foundation
of China; the National Basic Research of China under Grant No.
2010CB833004, PCSIRT under Grant No. IRT0964, the Construct Program
of the National Key Discipline, and Hunan Provincial Natural Science
Foundation of China 11JJ7001.

\end{acknowledgments}

\end{document}